\documentclass[runningheads]{llncs}
\usepackage[T1]{fontenc}
\usepackage{graphicx}
\usepackage{amsmath,amsfonts}
\usepackage{array}
\usepackage{booktabs}
\usepackage{enumitem}
\usepackage{subfig} 
\usepackage{textcomp}
\usepackage{url}
\usepackage{verbatim}
\usepackage[most]{tcolorbox}

\usepackage{xspace}
\usepackage{xcolor}
\usepackage{multirow}
\usepackage{pdflscape}
\usepackage{siunitx}
\usepackage{tabularx}
\usepackage{ltablex} 
\usepackage{algpseudocode}
\usepackage{mdframed}
\usepackage{longtable}
\usepackage{adjustbox}

\makeatother

\begin{document}


\title{Cost-Performance Analysis: A Comparative Study of CPU-Based Serverless and GPU-Based Training Architectures}
\titlerunning{Serverless vs GPU Training Architectures}

\author{Amine Barrak\inst{1} \and
Fabio Petrillo\inst{2} \and
Fehmi Jaafar\inst{3}}

\authorrunning{A. Barrak et al.}

\institute{
Oakland University, Rochester, MI, USA \\
\email{aminebarrak@oakland.edu} \and
École de technologie supérieure (ETS), Montréal, Canada \and
University of Quebec at Chicoutimi, Saguenay, Canada
}

\maketitle

\begin{abstract}
This paper presents a comparative evaluation of four serverless training frameworks: SPIRT, MLLess, LambdaML AllReduce, and ScatterReduce, alongside a GPU-based baseline, using CNN models on CIFAR-10. We assess each architecture across training time, cost, communication overhead, and accuracy under consistent experimental conditions. While GPU-based training achieves the fastest convergence and highest accuracy, serverless frameworks offer cost advantages for lightweight models. Optimizations such as gradient accumulation and in-database computation improve serverless performance. Our findings reveal key trade-offs and highlight the potential of GPU-backed serverless platforms for scalable distributed training.
\end{abstract}

\section{Introduction}
\label{sec:introduction}
Machine Learning (ML) continues transforming industries through increasingly sophisticated predictive capabilities. However, these advancements entail substantial computational demands and complex infrastructure management, especially when training moderately-sized models~\cite{you2018imagenet}. For instance, training ResNet-50 on ImageNet takes roughly 14 days using a single NVIDIA GPU~\cite{you2018imagenet}, prompting the adoption of distributed training methods to parallelize workloads, reducing training duration and resource bottlenecks~\cite{DistributedNN}.

Distributed ML architectures typically follow Parameter Server (PS) or Peer-to-Peer (P2P) paradigms~\cite{verbraeken2020survey}. PS centralizes gradient aggregation through a coordinator node, simplifying synchronization~\cite{li2013parameter}. Conversely, P2P decentralizes training by distributing parameters and computations across nodes, alleviating central coordination bottlenecks~\cite{vsajina2023peer}. Nevertheless, challenges remain, including resource inefficiencies, high operational costs, and system complexity.

The increasing computational demands of ML have driven a significant shift toward cloud-based training solutions. Cloud platforms offer scalable infrastructure, cost-effectiveness, and integrated support for parallel training \cite{mungoli2023scalable}. However, these benefits often come at the cost of inefficiencies such as over-provisioning and wasted resources. In fact, over-provisioning and always-on resources were estimated to account for \$26.6 billion in public cloud waste in 2021~\cite{Overprov70}.
A recent survey revealed that 41.1\% of data scientists find the use of cloud resources for ML training to be a significant challenge, highlighting the complexity of effectively managing such infrastructure~\cite{makinen2021needs}.

Serverless computing, characterized by on-demand resource allocation and seamless scaling, has emerged as a compelling alternative. It offers potential efficiency improvements by dynamically matching resource allocation with workload demands \cite{yu2022accelerating,mlless}. Recent studies validate the effectiveness of serverless computing frameworks in ML, particularly for smaller neural network architectures such as MobileNet and ResNet variants \cite{gyawali2023comparative,liu2022funcpipe,petrescu2023toward,barrak2023exploring,barrak2024incorporating}.

Several solutions utilizing serverless computing for distributed ML training have been proposed, adopting distinct architectural topologies such as Peer-to-Peer (P2P) \cite{barrak2023spirt}, Parameter Server (PS) \cite{P49,mlless}, ScatterReduce, and AllReduce \cite{P46}. A common challenge among these solutions is the stateless nature of serverless computing, requiring external communication channels to maintain state between functions \cite{P49,P54}. Techniques such as extending function runtime for state checkpointing \cite{P46} or enabling gradient accumulation to optimize parallel processing \cite{barrak2023spirt} have been explored to address these limitations.

Despite recent advancements, comprehensive comparisons of existing serverless ML frameworks remain limited, particularly against distributed GPU-based training setups. This paper addresses these gaps through a detailed analysis of end-to-end serverless architectures and an empirical evaluation comparing serverless CPU-based frameworks with distributed GPU-based baselines. To foster reproducibility and further research, we provide a full replication package \footnote{https://sites.google.com/view/pdcat-2025}.

The key contributions of this paper include:
\begin{enumerate}
    \item An analysis and comparison of existing serverless ML training architectures, highlighting their respective end-to-end frameworks and operational characteristics.

    \item Evaluation of communication overhead reduction techniques in serverless based training architectures, including in-database operations, update filtering, and communication pattern optimization.

    \item A detailed cost-performance analysis, identifying conditions under which serverless architectures offer substantial economic and performance advantages over GPU-based training.

\end{enumerate}

\section{Comparative Analysis of Training Workflows}
\label{sec:analysis}

\begin{figure}[t]
  \centering
  \includegraphics[width=\linewidth]{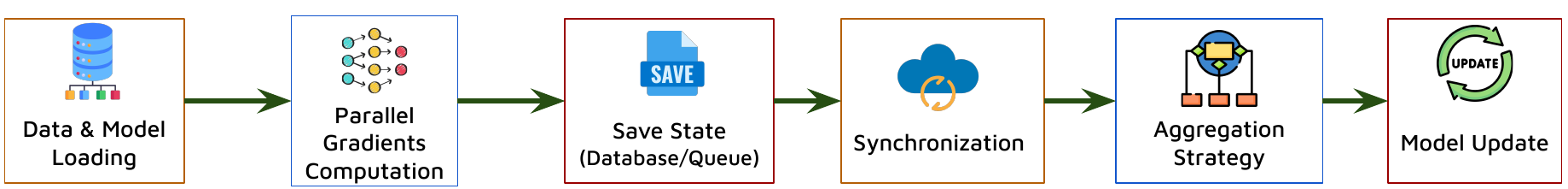}
  \caption{Abstract training flow shared by all evaluated architectures. Aggregation strategy and synchronization mechanisms vary per framework.}
  \label{fig:abstract_flow}
\end{figure}

Figure~\ref{fig:abstract_flow} illustrates a generalized training workflow in serverless environments. The process includes data fetching, model and dataset loading, gradient computation, saving intermediate state, synchronization, aggregation, and model update. Each framework instantiates this flow differently, reflecting trade-offs in communication overhead, synchronization complexity, and state handling.

SPIRT \cite{barrak2023spirt} decentralizes the entire workflow using a peer-to-peer strategy. Each worker independently loads its portion of data and model, computes gradients in parallel over minibatches, and saves these locally in its Redis instance. After averaging gradients internally, workers notify a synchronization queue and wait until all peers report completion. Once synchronized, workers retrieve each other’s averaged gradients, aggregate them again, and store the final result locally before applying the update. SPIRT uniquely performs both gradient averaging and model updates within the database, reducing communication with external storage and enhancing performance under stateless constraints.

MLLess \cite{mlless} adopts a significance-driven filtering strategy. Each worker loads data and model, computes gradients, and checks whether the change exceeds a predefined threshold. Only significant updates are stored in a shared database, and a message is pushed to other workers and a central supervisor via queues. The supervisor coordinates synchronization by instructing workers when to fetch updates. Workers then aggregate received gradients and update the model. This design reduces communication overhead but introduces a synchronization bottleneck through the central supervisor.

ScatterReduce \cite{P46} implements a distributed aggregation strategy by splitting gradients into chunks. Each worker computes gradients and retains one chunk while pushing the rest to shared storage. Workers are assigned specific chunks to aggregate across all peers, then upload the partial aggregates. After all chunks are aggregated, workers download the full set, reconstruct the complete gradient, and perform the model update. This strategy balances computation but incurs significant communication overhead, especially as the number of workers increases.

AllReduce \cite{P46} centralizes aggregation. After computing gradients, all workers push their updates to a shared database. A designated master worker performs the full aggregation and uploads the result. All workers then synchronize by fetching this final gradient and applying it to their local models. While simple and efficient for small setups, AllReduce suffers from a scalability bottleneck due to the master’s workload and centralized coordination.

GPU-Based Baseline uses multiple GPUs distributed across separate machines. Each GPU processes a different data batch, computes gradients locally, uploads them to a shared S3 bucket, then loads the gradients from other workers and performs local averaging before updating its model.




\begin{table*}[h!]
\centering
\caption{Comparative Analysis of ML Training Architectures: Overview of Key Computational Stages in Serverless and Distributed GPU-Based Frameworks}
\label{tab:communication_transposed}
\begin{adjustbox}{max width=12cm}
\begin{tabular}{|p{0.15\textwidth}|p{0.19\textwidth}|p{0.75\textwidth}|}
\hline
\textbf{Framework} & \textbf{Training Stage} & \textbf{Content of the Stage} \\ \hline

\multirow{4}{*}{SPIRT \cite{barrak2023spirt}}  
& Fetch Dataset & Each worker fetches its assigned minibatches. \\ \cline{2-3}
& \begin{tabular}[c]{@{}l@{}}Compute \\ Gradients\end{tabular}
  & Gradients are computed in parallel for each minibatch, sent to the local Redis database, and averaged within the database. \\ \cline{2-3}
& Synchronisation & The worker notifies a synchronization queue once gradients are ready, polls the queue until all peers complete, then retrieves averaged gradients from other workers, aggregates them, and stores the result locally. \\ \cline{2-3}
& Model Update & The final aggregated gradient is used to update the model. \\ \hline \hline

\multirow{4}{*}{MLLess \cite{mlless}}  
& Fetch Dataset & Each worker fetches a single minibatch for processing. \\ \cline{2-3}
& \begin{tabular}[c]{@{}l@{}}Compute \\ Gradients\end{tabular} & Gradients are computed for the minibatch and, if the change is significant, stored in a shared database with keys sent to peers via queues. \\ \cline{2-3}
& Synchronisation & Workers listen to their queues, collect update keys, wait for synchronization instructions from the supervisor, then fetch and aggregate the corresponding gradients. \\ \cline{2-3}
& Model Update & The aggregated gradients are used to update the model. \\ \hline \hline

\multirow{4}{*}{\begin{tabular}[c]{@{}l@{}}Scatter \\ Reduce \cite{P46}\end{tabular}}  
& Fetch Dataset & Each worker fetches a minibatch to process. \\ \cline{2-3}
& \begin{tabular}[c]{@{}l@{}}Compute \\ Gradients\end{tabular} & Gradients are computed and divided into chunks, with each chunk destined for a specific peer. Workers retain one chunk and send the rest to the database. \\ \cline{2-3}
& Synchronisation & Workers fetch chunks assigned to them, aggregate them, send the result back to the database, then retrieve and concatenate all aggregated chunks to form the full gradient. \\ \cline{2-3}
& Model Update & The full aggregated gradient is used to update the model. \\ \hline \hline

\multirow{4}{*}{\begin{tabular}[c]{@{}l@{}}All \\ Reduce \cite{P46}\end{tabular} }  
& Fetch Dataset & Each worker fetches a minibatch. \\ \cline{2-3}
& \begin{tabular}[c]{@{}l@{}}Compute \\ Gradients\end{tabular} & Gradients are computed for the minibatch and stored in a shared database. \\ \cline{2-3}
& Synchronisation & A designated master worker retrieves all gradients, performs aggregation, stores the result in the shared database, and the other workers fetch the aggregated gradient. \\ \cline{2-3}
& Model Update & Workers apply the aggregated gradient to update the model. \\ \hline \hline

\multirow{4}{*}{\begin{tabular}[c]{@{}l@{}}GPU-Based \\ (Distributed)\end{tabular}
}  
& Fetch Dataset & Each GPU loads its assigned batch of data and a local copy of the model. \\ \cline{2-3}
& \begin{tabular}[c]{@{}l@{}}Compute \\ Gradients\end{tabular} & Gradients are computed locally by each GPU. \\ \cline{2-3}
& Synchronisation & Each GPU uploads its gradients to a shared S3 bucket, retrieves others’ gradients, and performs local averaging. \\ \cline{2-3}
& Model Update & The locally averaged gradients are used to update the model. \\ \hline

\end{tabular}
\end{adjustbox}
\end{table*}

\section{Experimental Design \& Setup}
\label{sec:setup}

This section outlines the experimental design used to evaluate training performance, communication behavior, and final accuracy across multiple architectures: SPIRT, MLLess, ScatterReduce, AllReduce, and a distributed GPU-based baseline.

\subsection{Evaluation Metrics}
We conduct a fair and consistent comparison using the following metrics:

\textbf{Training Time and Cost per Epoch:} One epoch refers to a complete pass over the dataset. For serverless frameworks, we aggregate function execution time across all workers and compute cost using AWS Lambda pricing based on allocated memory and runtime duration. For the GPU setup, cost is derived from the total training time and the hourly price of the EC2 \texttt{g4dn.xlarge} instance.

\textbf{Communication Overhead:} Serverless training is stateless, requiring each function to load models and datasets upon invocation. We analyze the synchronization time and volume of data exchanged. SPIRT reduces communication overhead via in-database aggregation, while MLLess minimizes traffic using significant update filtering.

\textbf{Final Accuracy Evaluation:} Accuracy is recorded after each epoch to evaluate convergence behavior. We consider both final accuracy and the time required to achieve convergence.

\subsection{Dataset and Models}
The CIFAR-10 dataset \cite{krizhevsky2010convolutional} contains 60{,}000 $32\times32$ RGB images categorized into 10 classes. We evaluate the following models:

\noindent\textbf{MobileNet:} A lightweight architecture with approximately 4.2 million parameters, utilizing depth-wise separable convolutions for inference and training.

\noindent\textbf{ResNet-18:} A moderately deep network with 11.7 million parameters and skip connections, representing a lightweight yet representative architecture suitable for benchmarking scalability.

\subsection{Runtime Environment}
All experiments were conducted on AWS infrastructure:

\textbf{Serverless Setup:}  
AWS Lambda functions are deployed with stage-specific memory configurations. Core dependencies include PyTorch, NumPy, RedisAI, and \texttt{sshtunnel}. Lambda deployment is constrained to a 250MB unzipped package. SPIRT uses AWS Step Functions for orchestration, RedisAI hosted on EC2 for in-database operations, and RabbitMQ for worker synchronization. MLLess and LambdaML variants utilize Redis and either RabbitMQ or S3 as communication backends.

\textbf{GPU Baseline:}  
Training is distributed across multiple \texttt{g4dn.xlarge} EC2 instances, which are the cheapest GPU option available from AWS. Each instance includes an NVIDIA T4 GPU and 16 GB memory. GPUs process separate batches and synchronize via S3. Training settings match the serverless setups, and cost is based on on-demand pricing.

\textbf{Cost Estimation:}  
We use the AWS Pricing Calculator and official service rates to compute costs for compute time, storage interactions, orchestration steps, and message queues.

\section{Evaluation \& Results}
\label{sec:results}

This section presents a comprehensive evaluation of five training frameworks: SPIRT, MLLess, LambdaML AllReduce, LambdaML ScatterReduce, and a GPU baseline. The comparison covers training time, monetary cost, communication overhead, and model convergence.

\subsection{Training Time and Implied Cost}

\noindent \textbf{Motivation:}  
While GPU-based training is widely adopted due to high performance, it often results in over-provisioning and underutilization, especially in workloads with sporadic or parallelizable computations. Serverless computing offers a fine-grained billing model where users are charged only for the exact memory and compute time used per function. This subsection evaluates the cost-effectiveness of serverless architectures compared to GPU-based training when training CNN models at scale.

\noindent \textbf{Approach:}  
We evaluate four serverless training architectures: SPIRT, ScatterReduce, AllReduce, and MLLess, on two CNN models, MobileNet and ResNet-18, using the CIFAR-10 dataset with a batch size of 512. Each serverless variant runs 24 parallel Lambda function executions (batches) per worker node, across 4 total workers. For each architecture, we record:  
(i) the average duration per batch,  
(ii) the allocated memory per function, and  
(iii) the implied cost based on AWS Lambda’s x86 pricing.

AWS Lambda charges \$0.0000166667 per GB-second. The cost of a single function execution is:

\[
\text{Cost} = \text{Time (s)} \times \text{RAM (GB)} \times 0.0000166667
\]

\textit{Example:} For SPIRT running MobileNet, each function runs for 15.44 seconds with 2685 MB of memory (2.685 GB):  
\[
\text{Cost/function} = 15.44 \times 2.685 \times 0.0000166667 \approx 0.000689 \text{ USD}
\]
With 24 such functions per worker:  
\[
\text{Cost/worker} = 24 \times 0.000689 = 0.0165 \text{ USD}, \quad \text{Cost/4 workers} = 0.0660 \text{ USD}
\]

GPU cost is estimated using 4 \texttt{g4dn.xlarge} instances, each running for 92 seconds per epoch for MobileNet and 139 seconds for ResNet-18, with an on-demand pricing of \$0.526 per hour.

\noindent \textbf{Results:}  
Table~\ref{tab:summary_cost_24} summarizes total training time, peak memory usage, and implied cost per epoch across frameworks for both MobileNet and ResNet-18. 

For MobileNet, serverless frameworks, particularly ScatterReduce and AllReduce, were more cost-effective than the GPU baseline. This is because their execution time and memory requirements remained low enough that the serverless billing model remained efficient. 

However, with ResNet-18, the GPU baseline became cheaper. The deeper architecture increased both execution time and memory requirements for serverless functions. Since Lambda pricing multiplies time by allocated RAM, this led to higher costs. SPIRT and MLLess were especially impacted due to their design choices that maintain memory longer or involve heavier orchestration.

\begin{table}[h!]
\centering
\caption{Summary of Training Time, Peak RAM, and Cost per Epoch for MobileNet and ResNet-18}
\label{tab:summary_cost_24}
\begin{adjustbox}{max width=\textwidth}
\begin{tabular}{|p{2.5cm}|p{3.3cm}|p{1.5cm}|p{3.5cm}|p{1.4cm}|}
\hline
\textbf{Framework} & \textbf{Total Time (s)} & \textbf{Peak RAM (MB)} & \textbf{Cost/Worker (USD)} & \textbf{Total Cost (USD)} \\
\hline
\multicolumn{5}{|c|}{\textbf{MobileNet (CIFAR-10)}} \\
\hline
SPIRT              & 15.44 × 24 = 370.56     & 2685  & 24 × 0.000689 = 0.0165   & 0.0660 \\
ScatterReduce      & 14.343 × 24 = 344.23    & 2048  & 24 × 0.000442 = 0.0106   & 0.0422 \\
AllReduce          & 14.382 × 24 = 345.17    & 2048  & 24 × 0.000445 = 0.0107   & 0.0427 \\
MLLess             & 69.425 × 24 = 1666.20   & 3024  & 24 × 0.003496 = 0.0839   & 0.3356 \\
GPU (g4dn.xlarge)  & 92.00                   & N/A   & 0.01344                  & 0.0538 \\
\hline
\multicolumn{5}{|c|}{\textbf{ResNet-18 (CIFAR-10)}} \\
\hline
SPIRT              & 28.55 × 24 = 685.20     & 3200  & 24 × 0.001523 = 0.0365   & 0.1460 \\
ScatterReduce      & 27.17 × 24 = 652.08     & 2880  & 24 × 0.001302 = 0.0312   & 0.1249 \\
AllReduce          & 26.79 × 24 = 642.96     & 2986  & 24 × 0.001382 = 0.0332   & 0.1328 \\
MLLess             & 78.39 × 24 = 1881.36    & 3630  & 24 × 0.004737 = 0.1137   & 0.4548 \\
GPU (g4dn.xlarge)  & 139.00                  & N/A   & 0.0203                   & 0.0812 \\
\hline
\end{tabular}
\end{adjustbox}
\begin{flushleft}
\footnotesize
All results are based on training MobileNet and ResNet-18 over CIFAR-10 with a batch size of 512 using 4 workers.
\end{flushleft}
\end{table}

\begin{tcolorbox}[colback=gray!10!white, colframe=gray!50!black, title=\textbf{Findings}, fonttitle=\bfseries]
Serverless is more cost-effective for lightweight models like MobileNet. For deeper models like ResNet-18, GPU becomes cheaper due to lower sensitivity to memory and execution time.
\end{tcolorbox}

\subsection{Communication Overhead Reduction Techniques}
\label{sec:comm_overhead}

\noindent \textbf{Motivation:}
Serverless training frameworks suffer from communication overhead due to stateless operations and frequent synchronization. Reducing communication is critical for improving scalability and performance. We analyze three methods addressing this issue: significant update filtering (MLLess), in-database operations (SPIRT), and communication pattern tuning (LambdaML).

\noindent \textbf{Approach:}
Experiments were conducted using CIFAR-10 with MobileNet and ResNet variants. SPIRT was evaluated against a naive fetch-update-store baseline using Redis on ResNet-18. LambdaML was tested with MobileNet and ResNet-50 using 4 to 16 workers to compare ScatterReduce and AllReduce strategies. MLLess was extended to support MobileNet and evaluated using significant update filtering to delay parameter propagation based on gradient norms.

\begin{figure*}[h!]
    \centering

    \begin{minipage}[t]{0.63\textwidth}
        \centering
        \includegraphics[width=0.49\linewidth]{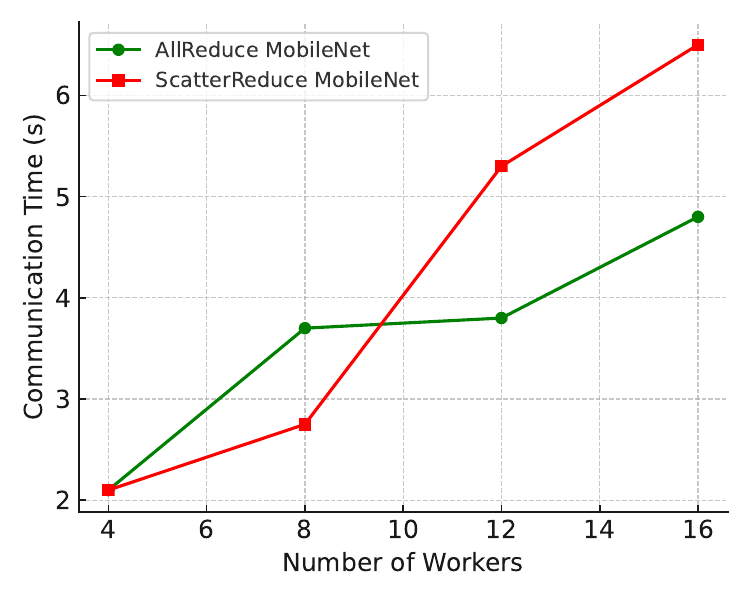}
        \includegraphics[width=0.49\linewidth]{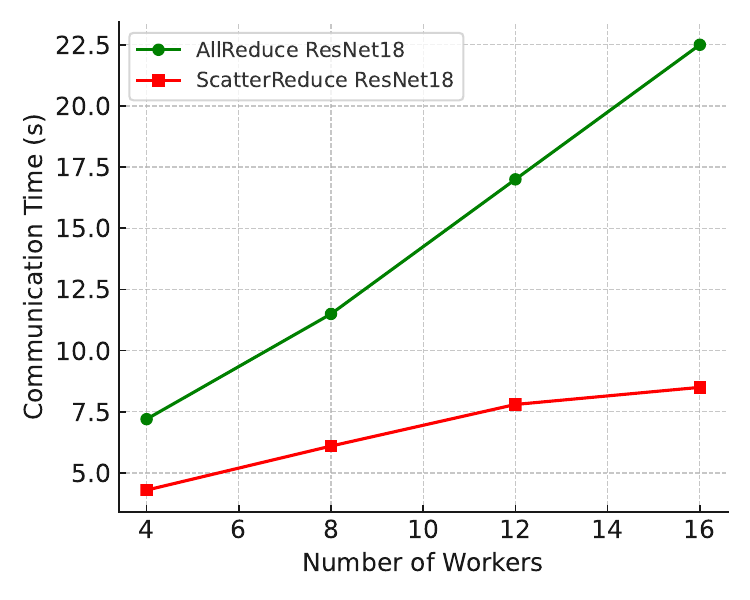}
        \caption{Communication time comparison between AllReduce and ScatterReduce under LambdaML using MobileNet and ResNet-18.}
        \label{fig:lambdaml_comm}
    \end{minipage}%
    \hfill
    \begin{minipage}[t]{0.35\textwidth}
        \centering
        \includegraphics[width=\linewidth]{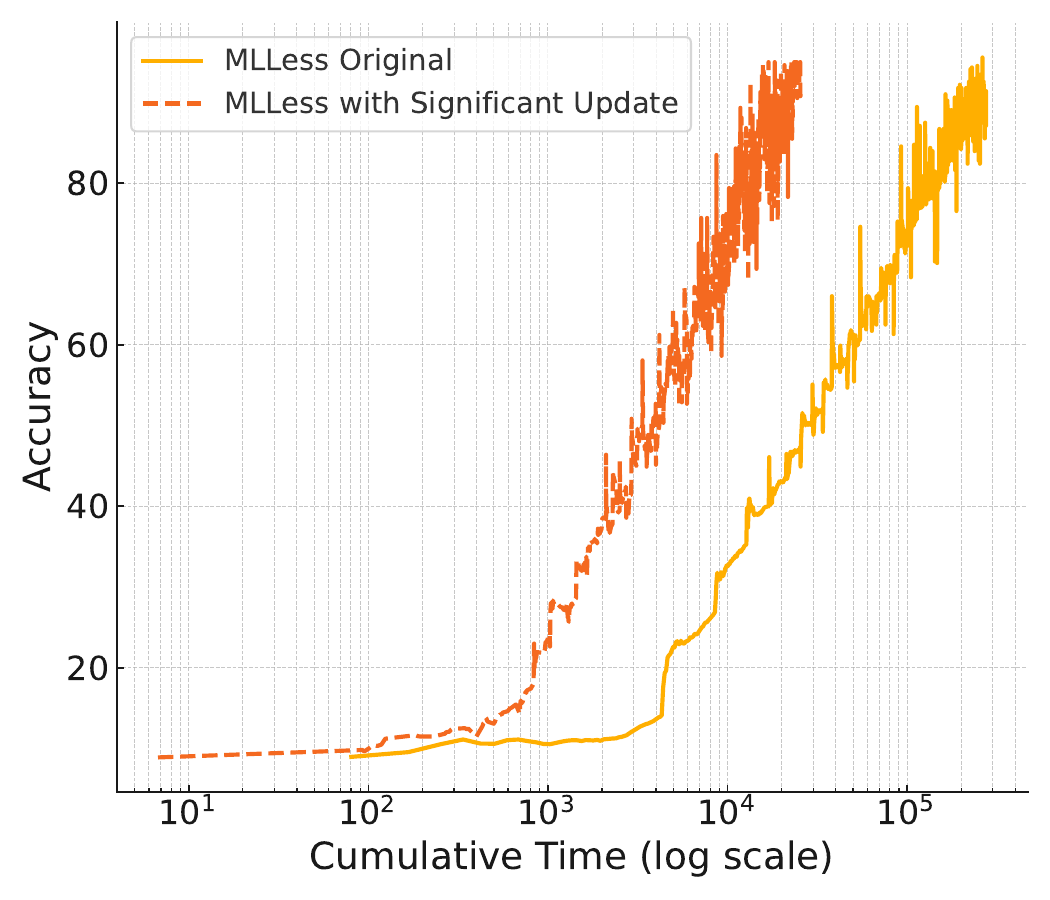}
        \caption{Communication overhead reduction using significant update filtering in MLLess with MobileNet.}
        \label{fig:mlless_comm}
    \end{minipage}

\end{figure*}

\noindent \textbf{Results:}
\noindent \textbf{MLLess} reduced convergence time from 113,379s to 8,667s using significant update filtering—a 13× improvement—while reducing communication by sending fewer updates.

\noindent \textbf{SPIRT} reduced gradient averaging time from 67.32s to 37.41s and model update time from 27.5s to 4.8s by operating directly within RedisAI. These operations avoided frequent external reads/writes compared to the naive baseline.

\noindent \textbf{LambdaML} results show that for the large ResNet-50 model, AllReduce scales poorly (up to 21.88s) compared to ScatterReduce (max 8.36s). For MobileNet, AllReduce performs better at higher worker counts (4.77s vs 6.47s with 16 workers).

\vspace{1em}
\begin{tcolorbox}[colback=gray!10, colframe=gray!80!black, title=\textbf{Findings}]
\small
Each method has strengths: MLLess reduces communication between workers by sharing only significant updates, SPIRT minimizes overhead through in-database operations, AllReduce handles larger models effectively with structured aggregation, while ScatterReduce can face worker bottlenecks as model size increases.
\end{tcolorbox}

\subsection{Performance Evaluation of Training Accuracy}
\label{sec:accuracy}

\noindent \textbf{Motivation:}
After evaluating training time and cost, it is essential to assess accuracy to understand the trade-offs between speed, cost, and model quality. This provides a balanced view of each framework’s effectiveness.

\noindent \textbf{Approach:}
We evaluated SPIRT, ScatterReduce, AllReduce, and MLLess using the MobileNet model on the CIFAR-10 dataset with a batch size of 512 per worker across 4 workers (global batch size 2{,}048). Each worker processed 24 full batches per epoch. In SPIRT and MLLess, these batches were pre-partitioned and scheduled for execution on each worker. In AllReduce and ScatterReduce, the dataset was split evenly, with each worker acting as a dataloader processing its 24 batches step by step. For the GPU-based baseline, the same global batch size of 2{,}048 was used on a single multi-GPU instance, with data parallelism distributing the 24 batches per device per epoch. Early stopping was applied to detect convergence in all setups.

\begin{figure}[h!]
 \centering
 \includegraphics[scale=0.36]{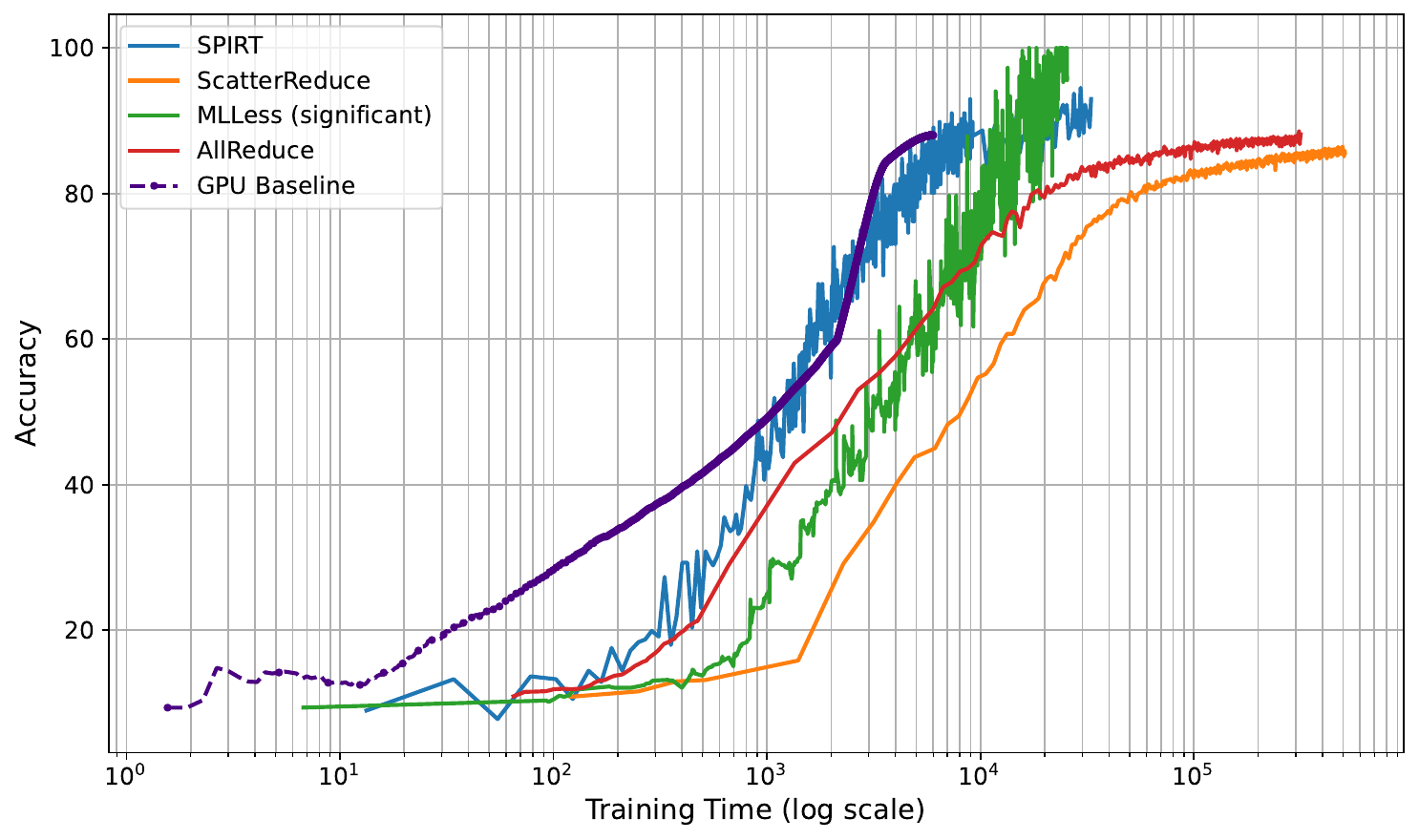}
\caption{Comparative accuracy evaluation of serverless training frameworks (log scale) with training time in seconds.}

 \label{fig:accuracy}
\end{figure}

\noindent \textbf{Results:}
\noindent \textbf{Results:}
Figure~\ref{fig:accuracy} shows the convergence behavior across all frameworks. The GPU-based baseline achieves 84.5\% accuracy in 70.33 minutes, demonstrating rapid convergence due to tightly coupled memory and compute. We also observe several fluctuations in SPIRT and MLLess curves, due to the effects of gradient accumulation and update scheduling. SPIRT reaches 83.2\% accuracy in 84.96 minutes, benefiting from in-database gradient accumulation that reduces communication overhead and accelerates training. MLLess converges more slowly, reaching 83.48\% after 189.68 minutes due to delayed updates and orchestration latency. ScatterReduce, although scalable, shows a slower and more gradual learning trajectory, converging at 82.1\% after 1652.49 minutes. AllReduce takes 1367.01 minutes to converge but achieves the highest accuracy at 85.05\%.

\begin{table}[h!]
  \caption{Convergence time and final accuracy (MobileNet, CIFAR-10).}
  \label{tab:acc}
  \centering
  \begin{tabular}{lcc}
    \toprule
    Framework & Time to 80\% (min) & Final Accuracy (\%) \\
    \midrule
    SPIRT            & 84.96  & 83.2 \\
    MLLess           & 189.68 & 83.48 \\
    ScatterReduce    & 1652.49 & 82.1 \\
    AllReduce        & 1367.01 & {85.05} \\
    GPU-based        & 70.33  & 84.5 \\
    \bottomrule
  \end{tabular}
\end{table}

\begin{tcolorbox}[colback=gray!10!white, colframe=gray!50!black, title=\textbf{Findings}, fonttitle=\bfseries]
GPU-based is fastest with strong accuracy. SPIRT offers the best trade-off between time and quality. MLLess is slower with similar performance. AllReduce achieves the highest accuracy but converges slowly. ScatterReduce is the slowest.
\end{tcolorbox}

\section{Discussion}
\label{sec:discussions}

This study reveals important trade-offs between cost, architectural complexity, and training efficiency across serverless and GPU-based frameworks.

\noindent \underline{\textbf{Cost–Performance and Framework Complexity:}}
Serverless computing provides flexibility and fine-grained billing, but the benefits vary across frameworks. MLLess and SPIRT incur higher costs due to orchestration overhead, particularly with deeper models. ScatterReduce and AllReduce offer a simpler and more cost-balanced approach for smaller models. The GPU-based baseline, while incurring higher hourly costs, achieves predictable performance and becomes more cost-effective for larger, compute-intensive workloads.


\noindent \underline{\textbf{Optimizations and Gradient Accumulation:}}
A key differentiator among the frameworks is their strategy for mitigating communication overhead. SPIRT employs in-database gradient accumulation to reduce synchronization frequency between workers. In contrast, MLLess utilizes a significance-driven filtering strategy, propagating only gradient updates that exceed a predefined threshold to minimize communication volume. While these optimizations improve training speed and convergence, their distinct architectural choices also directly influence training behavior, resulting in observable fluctuations in the learning curves of both frameworks.


\noindent \underline{\textbf{Serverless vs. GPU-Based Training:}}  
GPU-based training consistently provides faster convergence and higher accuracy with predictable performance , becoming more economical than serverless for heavier workloads due to its tight memory-compute integration and reduced overhead. In contrast, serverless frameworks remain appealing for lightweight or bursty tasks where their cost-per-usage and elastic scaling are prioritized over raw speed.

\noindent \underline{\textbf{The Synergy of Serverless and GPU Acceleration:}}
As this study demonstrates, a clear division exists in ML training architectures: serverless frameworks offer compelling cost advantages and elasticity for lightweight models , while GPU-based training provides the raw speed and power necessary for heavier workloads. The future of distributed training, however, may not lie in choosing between these paradigms, but in their synthesis. Emerging platforms from providers like Modal, Banana.dev, and RunPod already represent this next step, offering a new blend of serverless elasticity with GPU acceleration. By combining the pay-per-use cost model of serverless with the high-performance of GPUs could reshape the cost-performance trade-offs in distributed training.

\noindent \underline{\textbf{Threats to Validity:}}
The generalizability of our findings is primarily constrained by the experimental scope. The evaluation is confined to image classification on the CIFAR-10 dataset with two small CNNs (MobileNet and ResNet-18), and the observed cost-performance trade-offs may not extend to other data modalities or larger model architectures. Additionally, the results are specific to AWS infrastructure and its pricing model, which may not be transferable to other cloud providers. Finally, our cost model excludes the operational cost of database services, which were deemed negligible compared to compute-intensive operations, though slight variations in database access patterns could still introduce minor cost differences in real deployments.

\section{Conclusion}
\label{sec:conclusion}

This work presented a comparative evaluation of serverless and GPU-based training architectures across key metrics: training time, cost, communication overhead, and accuracy. Results show that GPU-based architecture consistently delivers the fastest convergence, achieving 84.5\% accuracy in 70.33 minutes. While serverless frameworks are significantly slowerIn terms of cost, serverless is more economical for lightweight models like MobileNet, but the GPU baseline becomes cheaper for heavier models like ResNet-18. Consequently, GPU training is preferred for workloads where speed and efficiency on complex tasks are paramount, while serverless is appealing for lightweight or bursty tasks where its elastic, pay-per-use model is most advantageous.

Future work points toward a synthesis of serverless and GPU paradigms to resolve the cost-performance trade-offs identified in this study.

\def\bibname{\section*{References}}
\bibliographystyle{splncs04}
\bibliography{references}

\end{document}